\date{November 29, 2004}

\documentclass{article}
\usepackage{latexsym}
\usepackage{amsfonts}
\usepackage{amsmath}
\usepackage{amsthm}
\usepackage{epsfig}
\usepackage{graphics}
\usepackage{verbatim}

\newtheorem{clm}{Claim}
\newtheorem{thm}{Theorem}
\newtheorem{defn}{Definition}
\newtheorem{cor}[thm]{Corollary}
\newtheorem{prop}[thm]{Proposition}
\newtheorem{lemma}[thm]{Lemma}

\newcommand{\E}{{\mathbb E}}

\newcommand{\R}{{\mathbb R}}

\newcommand{\dN} {{\widetilde{N}}}
\newcommand{\cN} {{\widehat{N}}}
\newcommand{\dn} {{\tilde{n}}}

\newcommand{\trans} {\mathsf{T}}
\newcommand{\var} {\mbox{\textup{var}}}
\newcommand{\prob} {{\bf P}}
\newcommand{\tee}{t}
\newcommand{\wpr} {{w^\prime}}

\newcommand{\xvec}{{\mathbf{x}}}
\newcommand{\yvec}{{\mathbf{y}}}
\newcommand{\zvec}{{\mathbf{z}}}
\newcommand{\Myz}{{M(\yvec,\zvec)}}

\newcommand{\rate} {{\mbox{rate}}}

\begin{document}
\begin{titlepage}
\title{Degree Distribution of Competition-Induced Preferential
Attachment Graphs}
\author{
N.~Berger$^{*}$ \and C.~Borgs$^{*}$ \and J.~T.~Chayes$^*$ \and
R.~M.~D'Souza\thanks{Microsoft Research, One Microsoft Way, Redmond
WA 98052, USA}
 \and
R.~D.~Kleinberg
\thanks{M.I.T. CSAIL, 77 Massachusetts Ave, Cambridge MA 02139, USA.
Supported by a Fannie and John Hertz Foundation Fellowship.}}

\maketitle \def\thepage {}
\begin{center}
{\it Dedicated to B. Bollob\'{a}s on the occasion of his 60th
birthday.}
\end{center}

\thispagestyle{empty}

\begin{abstract}

We introduce a family of one-dimensional geometric growth models,
constructed iteratively by locally optimizing the tradeoffs between
two competing metrics, and show that this family is equivalent to
a family of preferential attachment random graph models with upper
cutoffs.  This is the first explanation of how preferential attachment
can arise from a more basic underlying mechanism of local competition.
We rigorously determine the degree
distribution for the family of random graph models, showing that it
obeys a power law up to a finite threshold and decays exponentially
above this threshold.

We also rigorously analyze a generalized version of our graph
process, with two natural parameters, one corresponding to the
cutoff and the other a ``fertility'' parameter. We prove that the
general model has a power-law degree distribution up to a cutoff,
and establish monotonicity of the power as a function of the two
parameters.
Limiting cases
of the general model include the standard preferential attachment model
without cutoff and the uniform
attachment model.

\end{abstract}
\end{titlepage}
\pagenumbering {arabic}
\setcounter{page}{2}

\section{Introduction}
\subsection{Network growth models}

This paper is dedicated, with great affection and admiration, to
B\'{e}la Bollob\'{a}s on the occasion of his 60th birthday.  Two of
us (C.B. and J.T.C.) are privileged to count B\'{e}la among our
dearest friends. And all of us have been inspired by his pioneering
work on graph processes in general, and scale-free graphs in
particular.  We use the opportunity of this birthday volume to
provide complete proofs of results on a new graph model, first
announced in~\cite{BBCDK04}.

There is currently tremendous interest in understanding the
mathematical structure of networks -- especially as we discover
the pervasiveness of network structures in natural and engineered
systems. Much recent theoretical work has been motivated by
measurements of real-world networks, indicating they have certain
``scale-free'' properties, such as a power-law distribution of
degrees.  For the Internet graph, in particular, both the
graph of routers and the graph of autonomous systems (AS) seem to
obey power laws \cite{Falout3,GovinTangmu2000}.  However, these
observed power laws hold only for a limited range of degrees,
presumably due to physical constraints and the finite size of the
Internet.

Many random network growth models have been proposed which give rise
to power-law degree distributions.  Most of these models rely on a
small number of basic mechanisms, mainly preferential attachment%
\footnote{As Aldous \cite{Aldous-SCN} points out, proportional
attachment may be a more appropriate name, stressing the linear
dependence of the attractiveness on the degree.}
\cite{Price,BarabasiAlbert99} or copying \cite{KRRSTU}, extending
ideas known for many years \cite{Polya,Simon,Zipf,Yule} to a network
context.  Variants of the basic preferential attachment mechanism
have also been proposed, and some of these lead to changes in the
values of the exponents in the resulting power laws.  For extensive
reviews of work in this area, see Albert and Barab\'{a}si
\cite{BA-RMP02}, Dorogovtsev and Mendes \cite{DorogMendes02}, and
Newman \cite{MEJN-SIAM}; for a survey of the
relatively limited amount of mathematical work see
\cite{BolloRior02}.  Most of this work concerns network models
without reference to an underlying geometric space.  Nor do most of
these models allow for heterogeneity of nodes, or address physical
constraints on the capacity of the nodes.  Thus, while such models
may be quite appropriate for geometry-free networks, such as the web
graph, they do not seem to be ideally suited to the description of
other observed networks, {\em e.g.}, the Internet graph.

In this paper, instead of assuming preferential attachment, we show
that it can arise from a more basic underlying process, namely
competition between opposing forces.  The idea that power laws can
arise from competing effects, modeled as the solution of
optimization problems with complex objectives, was proposed
originally by Carlson and Doyle \cite{CD-HOT99}.  Their ``highly
optimized tolerance'' (HOT) framework has reliable design as a
primary objective.  Fabrikant, Koutsoupias and Papadimitriou (FKP)
\cite{FKP02} introduce an elegant network growth model with such a
mechanism, which they called ``heuristically optimized trade-offs''.
As in many growth models, the FKP network is grown one node at a
time, with each new node choosing a previous node to which it
connects.  However, in contrast to the standard preferential
attachment types of models, a key feature of the FKP model is the
underlying geometry.  The nodes are points chosen uniformly at
random from some region, for example a unit square in the plane. The
trade-off is between the geometric consideration that it is
desirable to connect to a nearby point, and a networking
consideration, that it is desirable to connect to a node that is
``central'' in the network as a graph.  Centrality is measured by
using, for example, the graph distance to the initial node. The
model has a tunable, but fixed, parameter, which determines the
relative weights given to the geometric distance and the graph
distance.

The suggestion that competition between two metrics could be an
alternative to preferential attachment for generating power-law
degree distributions represents an important paradigm shift.  Though
FKP introduced this paradigm for network growth, and FKP networks
have many interesting properties,
the resulting distribution is not a power law in the standard sense
\cite{BBBCR03}.  Instead the overwhelming majority of the nodes are
leaves (degree one), and a second substantial fraction heavily
connected ``stars'' (hubs), producing a node degree distribution
which
has clear bimodal features.%
\footnote{In simulations of the FKP model, this can be clearly
discerned by examining the probability distribution function (pdf);
for the system sizes amenable to simulations, it is less prominent
in the cumulative distribution function (cdf).}

Here, instead of directly producing power laws as a consequence of
competition between metrics, we show that such competition can give
rise to a preferential attachment mechanism, which in turn gives
rise to power laws.  Moreover, the power laws we generate have an
upper cutoff, which is more realistic in the context of many
applications.

\subsection{Overview of competition-induced preferential attachment}

We begin by formulating a general competition model for network
growth. Let $x_0, x_1, \dots, x_t$ be a sequence of random variables
with values in some space $\Lambda$.  We think of the points $x_0,
x_1, \dots, x_t$ arriving one at a time
according to some stochastic process.  For example, we typically
take $\Lambda$ to be a compact subset of $\mathbb R^d$, $x_0$ to be
a given point, say the origin, and $x_1, \dots, x_t$ to be i.i.d.
uniform on $\Lambda$.  The network at time $t$ will be represented
by a graph, $G(t)$, on $t+1$ vertices, labeled $0,1,\dots, t$, and
at each time step, the new node attaches to one or several nodes in
the existing network.  For simplicity, here we assume that each new
node connects to a single node, resulting in $G(t)$ being a tree.

Given $G(t-1)$, the new node, labeled $t$, attaches to that node $j$
in the existing network that minimizes a certain cost function
representing the trade-off of two competing effects, namely
connection or startup cost, and routing or performance cost.  The
connection cost is represented by a metric, $ g_{ij}(t)$, on $\{0,
\dots, t\}$ which depends on $x_0, \dots, x_{t}$, but not on the
current graph $G(t-1)$, while the routing cost is represented by a
function, $h_j(t-1)$, on the nodes which depends on the current
graph, but not on the physical locations $x_0,\dots,x_t$ of the
nodes $0,\dots,t$.  This leads to the cost function
\begin{equation}
c_{t} = {\min_j}\left[ \alpha g_{tj}(t)  + h_j(t-1) \right],
\label{cost}
\end{equation}
where $\alpha$ is a constant which determines the relative weighting
between connection and routing costs.  We think of the function
$h_j(t-1)$ as measuring the centrality of the node $j$; for
simplicity, we take it to be the hop distance along the graph
$G(t-1)$ from $j$ to the root $0$.

To simplify the analysis of the random graph process, we will assume
that nodes always choose to connect to a point which is closer to
the root, {\it i.e.}, they minimize the cost function
\begin{equation}
\tilde c_t = {\min_{j: \| x_j \| < \| x_{t} \|}} \left[ \alpha
g_{tj}(t)  + h_j(t-1) \right], \label{greedy-cost}
\end{equation}
where $\| \cdot \|$ is an appropriate norm.

In the original FKP model, $\Lambda$ is a compact subset of $\mathbb
R^2$, say the unit square, and the points $x_i$ are independently
uniformly distributed on $\Lambda$.  The cost function is of the
form \eqref{cost}, with $g_{ij} = d_{ij}$, the Euclidean metric
(modeling the cost of building the physical transmission line), and
$h_j(t)$ is the hop distance along the existing network $G(t)$ from
$j$ to the root.  A rigorous analysis of the degree distribution of
this two-dimensional model was given in \cite{BBBCR03}, and the
analogous one-dimensional problem was treated in \cite{KenyonSchab}.

Our model is defined as follows.

\begin{defn}[Border Toll Optimization Process]
Let $x_0=0$, and let $x_1,x_2,\dots$ be i.i.d., uniformly at random
in the unit interval $\Lambda = [0,1]$, and let $G(t)$ be the
following process: At $t=0$, $G(t)$ consists of a single vertex $0$,
the root. Let $h_j(t)$ be the hop distance to $0$ along $G(t)$, and
let $g_{ij}(t) = n_{ij}(t)$ be the number of existing nodes between
$x_i$ and $x_j$ at time $t$, which we refer to as the {\em jump
cost} of $i$ connecting to $j$.  Given $G(t-1)$ at time $t-1$, a new
vertex, labeled $t$, attaches to the node $j$ which minimizes the
cost function \eqref{greedy-cost}. Furthermore, if there are several
nodes $j$ that minimize this cost function and satisfy the
constraint, we choose the one whose position $x_j$ is nearest to
$x_t$.  The process so defined is called the {\em border toll
optimization process (BTOP)}.
\end{defn}

As in the FKP model, the routing cost is just the hop distance to
the root along the existing network.  However, in our model the
connection cost metric measures the number of ``borders'' between
two nodes: hence the name BTOP.  Note the correspondence to the
Internet, where the principal connection cost is related to the
number of AS domains crossed -- representing, {\em e.g.}, the
overhead associated with BGP, monetary costs of peering agreements,
etc.  In order to facilitate a rigorous analysis of our model, we
took the simpler cost function \eqref{greedy-cost}, so that the new
node always attaches to a node to its left.

It is interesting to note that the ratio of the BTOP connection cost
metric to that of the one-dimensional FKP model is just the local
density of nodes: $n_{ij}/d_{ij} = \rho_{ij}$.  Thus the
transformation between the two models is equivalent to replacing the
constant parameter $\alpha$ in the FKP model with a variable
parameter $\alpha_{ij} = \alpha\rho_{ij}$ which changes as the
network evolves in time.  That $\alpha_{ij}$ is proportional to the
local density of nodes in the network reflects a model with an
increase in cost for local resources that are scarce or in high
demand.  Alternatively, it can be thought of as reflecting the
economic advantages of being first to market.

Somewhat surprisingly, the BTOP is equivalent to a special case of
the following process, which closely parallels the preferential
attachment model and makes no reference to any underlying geometry.

\begin{defn}[Generalized Preferential Attachment with Fertility and Aging]
\label{def:cipa} Let $A_1,A_2$ be two positive integer-valued
parameters.  Let $G(t)$ be the following Markov process, whose
states are finite rooted trees in which each node is labeled either
{\em fertile} or {\em infertile}. At time $t=0$, $G(t)$ consists of
a single fertile vertex.  Given the graph at time $t$, the new graph
is formed in two steps: first, a new vertex, labeled $t+1$ and
initialized as infertile, connects to an old vertex $j$ with
probability zero if $j$ is infertile, and with probability
\begin{equation}
\label{cipa-pr} Pr(t+1 \rightarrow j) =
\frac{\min\{d_j(t),A_2\}}{W(t)}
\end{equation}
if $j$ is fertile.  Here, $d_j(t)$ is equal to $1$ plus the
out-degree of $j$, and $W(t)=\sum'_j \min\{d_j(t),A_2\}$ with the
sum running over fertile vertices only.  We refer to vertex $t+1$ as
a child of $j$. If after the first step, $j$ has more than $A_1-1$
infertile children, one of them, chosen uniformly at random, becomes
fertile. The process so defined is called a {\em generalized
preferential attachment process with fertility threshold $A_1$ and
aging threshold $A_2$.}

The special case $A_1=A_2$ is called the {\em competition-induced
preferential attachment process} with parameter $A_1$.
\end{defn}

The last definition is motivated by the following theorem, to be
proved in Section~\ref{sec:themodel}.  To state the theorem, we
define a graph process as a random sequence of graphs
$G(0),G(1),G(2),\ldots$ on the vertex sets $\{0\}, \{0,1\},
\{0,1,2\},\ldots,$ respectively.

\begin{thm}  \label{thm:equiv}
As a graph process, the border toll optimization process has the
same distribution as the competition-induced preferential attachment
process with parameter $A = \lceil \alpha^{-1} \rceil$.
\end{thm}

Certain other limiting cases of the generalized preferential
attachment process are worth noting. If $A_1 = 1$ and $A_2 =
\infty$, we recover the standard  model of preferential attachment
as considered in \cite{Price,BarabasiAlbert99}. If $A_1 = 1$ and
$A_2$ is finite, the model is equivalent to the standard model of
preferential attachment with a cutoff. On the other hand, if $A_1 =
A_2 = 1$, we get a uniform attachment model.

The degree distribution of our random trees is characterized by the
following theorem, which asserts that almost surely (a.s.) the
fraction of vertices having degree $k$ converges to a specified
limit $q_k$, and moreover that this limit obeys a power law for $k <
A_2$, and decays exponentially above $A_2$.

\begin{thm} \label{thm:main-thm}
Let $A_1$, $A_2$ be positive integers. Consider the generalized
preferential attachment process with fertility parameter $A_1$ and
aging parameter $A_2$.  Let $N_0(t)$ be the number of infertile
vertices at time $t$, and let $N_k(t)$ be the number of fertile
vertices with $k-1$ children at time $t$, $k\geq 1$.  Then:
\begin{enumerate}
\item \label{mthm:convergence}
There are numbers $q_k\in [0,1]$ such that, for all $k\geq 0$
\begin{equation}
\label{mthm:empconv} \frac{N_k(t)}{t+1}\to q_k\quad\text{a.s., as
}t\to\infty.
\end{equation}
\item \label{mthm:qk-formula}
There exists a number $w =w(A_1,A_2)\in [0,2]$ such that the $q_k$
are determined by the following equations:
\begin{eqnarray}
\label{mthm:1} q_i & = &
\left(\prod_{k=2}^i\frac{k-1}{k+w}\right)q_1
\quad\text{if}\quad 1 \leq i \leq A_2, \\
\label{mthm:2} q_i & = &
\left(\frac{A_2}{A_2+w}\right)^{i-A_2}q_{A_2} \quad\text{if}\quad i>
A_2
\end{eqnarray}
\[
1 = \sum_{i=0}^\infty q_i , \qquad\text{and}\qquad q_0  =
\sum_{i=1}^\infty q_i\min\{i-1,A_1-1\}.
\]
\item \label{mthm:powerlaw}
There are positive constants $c_1$ and $C_1$, independent of $A_1$
and $A_2$, such that
\begin{equation} \label{eqn:mainthm_powerlaw}
c_1 k^{-(w+1)} < q_k/q_1 < C_1 k^{-(w+1)}
\end{equation}
for $1 \le k \le A_2$.
\item \label{mthm:monotonicity}
If $A_1=A_2$, the parameter $w$ is equal to $1$, and for general
$A_1$ and $A_2$, $w$ decreases with increasing $A_1$, and increases
with increasing $A_2$.
\end{enumerate}
\end{thm}

Equation (\ref{eqn:mainthm_powerlaw}) clearly defines a power-law
degree distribution with exponent $\gamma = w+1$ for $k \leq A_2$.
Note that for measurements of the Internet the value of the exponent
for the power law is $\gamma \approx 2$.  In our border toll
optimization model, where $A_1 = A_2$, we recover $\gamma=2$.

The convergence claim of Theorem~\ref{thm:main-thm} is proved using
a novel method which we believe is one of the main technical
contributions of this work.  For preferential attachment models
which have been analyzed in the
past~\cite{ACL02,BBCR03,BRST01,CF01}, the convergence was
established using the Azuma-Hoeffding martingale inequality.  To
establish the bounded-differences hypothesis required by that
inequality, those proofs employed a clever coupling of the random
decisions made by the various edges, such that the decisions made by
an edge $e$ only influence the decisions of subsequent edges which
choose to imitate $e$'s choices.  A consequence of this coupling is
that if $e$ made a different decision, it would alter the degrees of
only finitely many vertices.  This in turn allows the required
bounded-differences hypothesis to be established.  No such approach
is available for our models, because the coupling fails.  The random
decisions made by an edge $e$ may influence the time at which some
node $v$ crosses the fertility or aging threshold, which thereby
exerts a subtle influence on the decisions of {\em every} future
edge, not only those which choose to imitate $e$.

Instead we introduce a new approach based on the second-moment
method. The argument establishing the requisite second-moment upper
bound is quite subtle; it depends on a computation involving the
eigenvalues of a matrix describing the evolution of the degree
sequence in a continuous-time version of the model.

\section{Equivalence of the Two Models}
\label{sec:themodel}

\subsection{Basic properties of the border toll optimization
process}

In this section we will turn to the BTOP defined in the
introduction, establishing some basic properties which will enable
us to prove that it is equivalent to the competition-induced
preferential attachment model. In order to avoid complications we
exclude the case that some of the $x_i$'s are identical, an event
that has probability zero.  We say that $j\in\{0,1\dots,t\}$ lies to
the right of $i\in\{0,1\dots,t\}$ if $x_i<x_j$, and  we say that $j$
lies directly to the right of $i$ if $x_i<x_j$ but there is no
$k\in\{1,\dots,t\}$ such that $x_i<x_k<x_j$.  In a similar way, we
say that $j$ is the first vertex with a certain property to the
right of $i$ if $j$ has that property and there exists no
$k\in\{1,\dots,t\}$ such that $x_i<x_k<x_j$ and $k$ has the property
in question.  Similar notions apply with ``left'' in place of
``right''.

\begin{defn}
A vertex $i$ is called {\em fertile at time $t$} if a hypothetical
new point arriving at time $t+1$ and landing directly to the right
of $x_i$ would attach itself to the node $i$.  Otherwise $i$ is
called {\em infertile} at time $t$.
\end{defn}

\begin{figure}[bct]
\vskip-1cm {\hfill
\resizebox{4.2in}{!}{\includegraphics{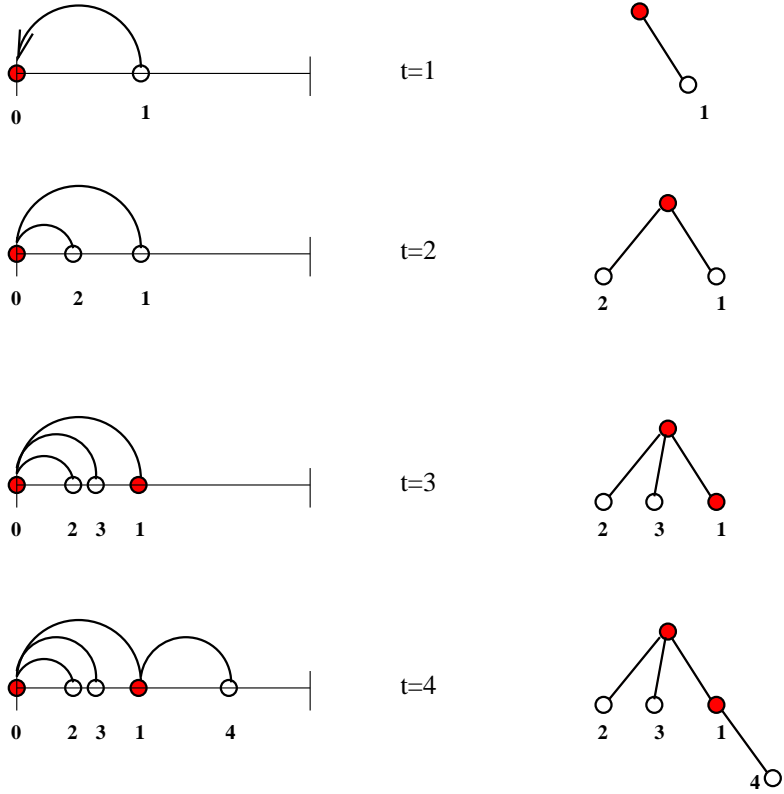}}\hfill}
\vspace{-0.2in} \caption{A sample instance of BTOP for $\alpha=1/3,
A=3$, showing the process on the unit interval (on the left), and
the resulting tree (on the right). Fertile vertices are shaded,
infertile ones are not. Note that vertex $1$ became fertile at
$t=3$.} \label{fig:tree-tA+1}
\end{figure}

This definition  is illustrated in Fig.~\ref{fig:tree-tA+1}.

\begin{lemma}
\label{lem:fertility} Let $0<\alpha<\infty$, let $A=\lceil
\alpha^{-1}\rceil$, and let $0<t<\infty$.  Then

i) The node $0$ is fertile at time $t$.

ii) Let $i$ be fertile at time $t$.  If $i$ is the rightmost fertile
vertex at time $t$ (case 1), let $\ell$ be the number of infertile
vertices to the right of $i$.  Otherwise (case 2), let $j$ be the
next fertile vertex to the right of $i$, and let $\ell=n_{ij}(t)$.
Then $0\leq \ell\leq A-1$, and the $\ell$ infertile vertices located
directly to the right of $i$ are children of $i$. In case 2, if
$h_{j}>h_i$, then $j$ is a fertile child of $i$ and $\ell=A-1$. As a
consequence, the hop count between two consecutive fertile vertices
never increases by more than $1$ as we move to the right, and if it
increases by $1$, there are $A-1$ infertile vertices between the two
fertile ones.

iii) Assume that the new vertex at time $t+1$ lands between two
consecutive fertile vertices $i$ and $j$, and let $\ell=n_{ij}(t)$.
Then $t+1$ becomes a child of $i$.  If $\ell+1<A$, the new vertex is
infertile at time $t+1$, and the fertility of all old vertices is
unchanged.  If $\ell+1=A$ and the new vertex lies directly to the
left of $j$, the new vertex is fertile at time $t+1$ and the
fertility of the old vertices is unchanged.  If $\ell+1=A$ and the
new vertex does not lie directly to the left of $j$, the new vertex
is infertile at time $t+1$, the vertex directly to the left of $j$
becomes fertile, and the fertility of all other vertices is
unchanged.

iv) If $t+1$ lands to the right of the rightmost fertile vertex at
time $t$, the statements in iii) hold with $j$ replaced by the right
endpoint of the interval $[0,1]$, and $n_{ij}(t)$ replaced by the
number of vertices to the right of $i$.

v) If $i$ is fertile at time $t$, it is still fertile at time $t+1$.

vi) If $i$ has $k$ children at time $t$, the $\ell=\min\{A-1,k\}$
leftmost of them are infertile at time $t$, and any others are
fertile.
\end{lemma}

\begin{proof}
The proof is straightforward but lengthy.  We include the details of
the argument here for completeness.

Statement i) is trivial, statement v) follows immediately from iii)
and iv), and vi) follows immediately from ii). So we are left with
ii)
--- iv).  We proceed by induction on $t$.  If ii) holds at time $t$,
and iii) and iv) hold for a new vertex arriving at time $t+1$, ii)
clearly also holds at time $t+1$.  We therefore only have to prove
that ii) at time $t$ implies iii) and iv) for a new vertex arriving
at time $t+1$.

Assume thus that ii) holds at time $t$. At time $t+1$, a new vertex
arrives, and falls directly to the right of some vertex $k$. Let $i$
be the nearest vertex to the left of $k$ that was fertile at time
$t$ (if $k$ is fertile at time $t$, we set $i=k$) and let $j$ be the
nearest vertex to the right of $i$ that was fertile at time $t$ (we
assume for the moment that $i$ is not the rightmost fertile vertex
at time $t$), let $\ell$ be the number of vertices between $i$ and
$j$ at time $t$.

Let us first prove that the vertex $t+1$ connects to $i$. If $i=k$,
this is obvious, since $i$ is fertile at time $t$. We may therefore
assume that $k\neq i$.  For the new vertex $t+1$, the cost of
connecting to the vertex $i$ is then equal to $\alpha
(n_{ik}(t)+1)$. Let us first compare this cost to the cost of
connecting to a fertile vertex $i'$ to the left of $i$. Let
$i_0=i'$, let $i_s=i$, and let $i_1,\dots,i_{s-1}$ be the fertile
vertices between $i'$ and $i$, ordered from left to right.  If
$h_{i_{m-1}}<h_{i_m}$, we use the inductive assumption ii) to
conclude that
 the number of infertile vertices between $i_{m-1}$ and $i_m$ is
equal to $A-1$, and $h_{i_{m-1}}=h_{i_m}-1$.  A decrease of $q$ in
the hop cost is therefore accompanied by an increase in the jump
cost of at least $\alpha A q\geq q$.  As a consequence, it never
pays to connect to a fertile vertex $i'$ to the left of $i$.  The
cost of connecting to an infertile vertex to the left of $i$ is even
higher, since the hop count of an infertile vertex is at best equal
to the hop count of the next fertile vertex to the right.  We
therefore only have to consider the connection cost to some of the
infertile children of $i$. But again, the hop count is worse by $1$
when compared to the hop count of $i$, and the jump cost is at best
reduced by $(A-1)\alpha<1$, proving that the cost of connecting to
$i$ is minimal.

To discuss the fertility of the vertices in the graph $G(t+1)$, we
need to consider the arrival of a second vertex, labeled $t+2$.  If
$t+2$ falls to the left of $t+1$, it will face an optimization
problem that has not been changed by the arrival of the vertex
$t+1$, implying that the fertility of the vertices to the left of
$t+1$ is unchanged.  If $t+2$ falls to the right of $j$, the cost of
connecting to $j$ or one of the vertices to the right of $j$ is the
same as before, and the cost of connecting to a vertex to the left
of $j$ is at best equal (the cost of connecting to any vertex to the
left of $t+1$ is in fact higher, due to the additional cost of
jumping over the vertex $t+1$).  Therefore, the vertex $t+2$ will
still prefer to connect to either $j$ or one of the vertices to the
right of $j$, implying that the fertility of the vertices to the
right of $j$ has not changed at all.  We therefore are left with
analyzing the case where $t+2$ falls between $t+1$ and $j$. Again,
the vertex $t+2$ will prefer $i$ over any vertex to the left of $i$
(the cost analysis is the same as the one used for $t+1$ above), so
we just have to compare the costs of connecting to the different
vertices between $i$ and $j$. If $\ell+1<A$, this will again imply
that $t+2$ connect to $i$; but if $\ell+1=A$, the vertex $t+2$ will
only connect to $i$ if it does not fall to the right of the
rightmost of the now $\ell+1$ vertices between $i$ and $j$.  If it
falls to the right of this vertex, it will be as expensive to
connect to the rightmost of the now $\ell+1$ vertices between $i$
and $j$ as it is to connect to $i$.  Recalling out convention of
connecting to the nearest vertex to the left if there is a tie in
costs, this proves that now $t+2$ connects to the rightmost vertex
between $i$ and $j$, implying that this vertex is fertile.

The above considerations prove the fertility statements in iii), and
thus completes the proof of iii). The case where $i$ is the
rightmost fertile vertex at time $t$ is similar (in fact, it is
slightly easier since it involves fewer cases), and leads to the
proof of iv).  This completes the proof of
Lemma~\ref{lem:fertility}.
\end{proof}

\subsection{Proof of Theorem~\ref{thm:equiv}}

In the BTOP, note that our cost function
\begin{equation} \label{eqn:costfunc-repeat}
{\rm min_j}\left[ \alpha n_{tj}(t) + h_j(t-1) \right],
\end{equation}
and hence the graph $G(t)$, only depends on the order of the
vertices $x_0,\dots,x_t$, and not on their actual positions in the
interval $[0,1]$.  Let $\vec\pi(t)$ be the permutation of
$\{0,1,\dots,t\}$ which orders the vertices $x_0,\dots,x_t$ from
left to right, so that
\begin{equation}
x_0=x_{\pi_0(t)}<x_{\pi_1(t)}<\dots<x_{\pi_t(t)}.
\end{equation}
(Recall that the vertices $x_0,x_1,\dots,x_t$ are pairwise distinct
with probability one.)  Note that $\vec\pi(t)$ and $\vec\pi(t+1)$
are related as follows: there exists $i_0 \in \{1,2,\ldots,t+1\}$
such that
\begin{equation} \label{eqn:pi-transition}
\pi_i(t+1)=
\begin{cases}
\pi_i(t)      &\text{if}\quad i < i_0\\
t +1          &\text{if}\quad i=i_0\\
\pi_{i-1}(t)&\text{if}\quad i>i_0.
\end{cases}
\end{equation}
Informally, the permutation $\vec\pi(t+1)$ is obtained by inserting
the new element $t+1$ into the permutation $\vec\pi(t)$ in a random
position $i_0$, where $x_{\pi_{i_0}(t)}$ is the left endpoint of the
subinterval of $(0,1)$ into which $x_{t+1}$ falls. The distribution
of the random variable $i_0$ may be deduced as follows. Since $x_0 =
0$ and $x_1,x_2,\ldots,x_t$ are i.i.d., we know that, for all $t$,
the permutation $\vec\pi(t)$ is uniformly distributed among
permutations of $\{0,1,\ldots,t\}$ which fix the element $0$.  This
means that, conditioned on a given such permutation $\vec\pi(t)$,
the permutation $\vec\pi(t+1)$ is uniformly distributed among all
permutations related to $\vec\pi(t)$ by the transformation
(\ref{eqn:pi-transition}).  In other words, $i_0$ is uniformly
distributed in the set $\{1,2,\ldots,t+1\}$.

With the help of Lemma~\ref{lem:fertility}, we now easily derive a
description of the graph $G(t)$ which does not involve any
optimization problem. To this end, let us consider a vertex $i$ with
$\ell$ infertile children at time $t$.  If a new vertex falls into
the interval directly to the right of $i$, or into one of the
intervals directly to the right of an infertile child of $i$, it
will connect to the vertex $i$.  Since there is a total of $t+1$
intervals at time $t$, the probability that a vertex $i$ with $\ell$
infertile children grows an offspring is $(\ell+1)/(t+1)$. By
Lemma~\ref{lem:fertility} (vi), this number is equal to
$\min\{A,k_i\}/(t+1)$, where $k_i-1$ is the number of children of
$i$. Note that fertile children do not contribute to this
probability, since vertices falling into an interval directly to the
right of a fertile child will connect to the child, not the parent.

Assume now that $i$ did get a new offspring, and that it had $A-1$
infertile children at time $t$.  Then the new vertex is  either born
fertile, or makes one of its infertile siblings fertile.
 Using the principle of deferred decisions,
 we may assume that with probability $1/A$ the new vertex becomes
 fertile, and with probability $(A-1)/A$ an old one, chosen
uniformly at random among the $A-1$ candidates, becomes fertile.

We thus have shown that the solution $G(t)$ of the optimization
problem \eqref{eqn:costfunc-repeat} can alternatively be described
by the competition-induced preferential attachment model with
parameter $A$.

\section{Convergence of the Degree Distribution}
\label{sec:a1eqa2}

\subsection{Overview}

To characterize the behavior of the degree distribution, we will
derive a recursion which governs the evolution of the vector
$\vec{N}(t)$, whose components are the number of vertices of each
degree, at the time when there are $\tee$ nodes in the network.  The
conditional expectation of $\vec{N}(t+1)$ is given by an evolution
equation of the form
\[
\E\left(\vec{N}(t+1) - \vec{N}(t) \;|\; \vec{N}(t)\right) =  M(t)
\vec{N}(t),
\]
where $M(t)$ depends on $\tee$ through the random variable $W(t)$
introduced in Definition~\ref{def:cipa}. Due to the randomness of
the coefficient matrix $M(t)$, the analysis of this evolution
equation is not straightforward.  We avoid this problem by
introducing a
continuous-time process, with time parameter $\tau$, which is
equivalent to the original discrete-time process up to a (random)
reparametrization of the time coordinate.  The evolution equation
for the conditional expectations in the continuous-time process
involves a coefficient matrix $M$ that is not random and does not
depend on $\tau$.  We will first prove that the {\em expected}
degree distribution in the continuous-time model converges to a
scalar multiple of the eigenvector $\hat{p}$ of $M$ associated with
the largest eigenvalue $w$.  This is followed by the much more
difficult proof that the {\em empirical} degree distribution
converges a.s. to the same limit.  Finally, we translate this
continuous-time result into a rigorous convergence result for the
original discrete-time system.

\subsection{Notation}\label{subseq:notation}

Let $A$ be any integer greater than or equal to $\max(A_1,A_2)$. Let
$N_0(\tee)$ be the number of infertile vertices at (discrete) time
$\tee$, and, for $k\geq 1$, let $N_k(\tee)$ be the number of fertile
vertices with $k-1$ children at time $\tee$.  Let
$\dN_A(\tee)=N_{\geq A}(\tee)=\sum_{k\geq A}N_k(\tee)$, and
$\dN_k(\tee)=N_k(\tee)$ if $k<A$.  The combined attractiveness of
all vertices is denoted by $W(\tee) = \sum_{k=1}^{A} \min\{k,A_2\}
\dN_k(\tee)$.  Let $n_k(\tee) = \frac{1}{\tee+1} N_k(\tee)$ and
$\dn_k(\tee) = \frac{1}{\tee+1} \dN_k(\tee)$.  Finally, the vectors
$( \dN_k(\tee) )_{k=1}^A$ and $( \dn_k(\tee) )_{k=1}^A$ are denoted
by $\dN(\tee)$ and $\dn(\tee)$ respectively.  Note that the index
$k$ runs from $1$ to $A$, not $0$ to $A$.

\subsection{Evolution of the expected value}
\label{subseq:evolexp}

From the definition of the generalized preferential attachment
model, it is easy to derive the probabilities for the various
alternatives which may happen upon the arrival of the $(\tee+1)$-st
node:
\begin{itemize}
\item  With probability $A_2 \dN_A(\tee)/W(\tee)$, it attaches to a node
of degree $\geq A$.  This increments $\dN_1$, and leaves $\dN_A$ and
all $\dN_j$ with $1<j<A$ unchanged.
\item  With probability $\min(A_2,k) \dN_k(\tee)/W(\tee)$, it attaches
to
a node of degree $k$, where $1 \leq k < A$. This increments
$\dN_{k+1}$, decrements $\dN_k$, increments $\dN_0$ or $\dN_1$
depending on whether $k < A_1$ or $k \geq A_1$, and leaves all other
$\dN_j$ with $j<A$ unchanged.
\end{itemize}
It follows that the discrete-time process $(\dN_k(\tee))_{k=0}^{A}$
at time $\tee$ is equivalent to the state of the following
continuous-time stochastic process $(\cN_k(\tau))_{k=0}^{A}$ at the
random stopping time $\tau = \tau_t$ of the $t$-th event.
\begin{itemize}
\item   With rate $A_2 \cN_A(\tau)$, $\cN_1$ increases by $1$.
\item   For every $0<k<A$, with rate
$\cN_k(\tau)\min(k,A_2)$, the following happens:
\[
\cN_k\to \cN_k-1\ \ \  ; \ \ \ \cN_{k+1}\to \cN_{k+1}+1\ \ \   ; \ \
\ \cN_{g(k)} \to \cN_{g(k)}+1
\]
where $g(k)=0$ for $k<A_1$ and $g(k)=1$ otherwise.
\end{itemize}
Note that the above rules need to be modified if $A_1=1$.  Here the
birth of a child of a degree-one vertex does not change the net
number of fertile degree-one vertices, $N_1$.

Let $M$ be the following $A\times A$ matrix:
\begin{equation} \label{eqn:M}
M_{i,j}=\left\{
\begin{array}{ll}
-1 & \mbox{if} \ i=j=1<A_1\\
-\min(j,A_2) & \mbox{if } \ 2\leq i=j\leq A-1 \\
\min(j,A_2) & \mbox{if } \ 2\leq i=j+1\leq A \\
\min(j,A_2) & \mbox{if } \ i=1 \mbox{ and } j\geq \max(A_1,2) \\
0 & \mbox{otherwise}.
\end{array}
\right.
\end{equation}
Then,
for every $\tau>\sigma$, the  conditional expectation of the vector
$\cN(\tau)=(\cN_k(\tau))_{k=1}^{A}$ is given by
\begin{equation}\label{eq:exval}
\E\left(\cN(\tau) \;|\;
\cN(\sigma)\right)=e^{(\tau-\sigma)M}\cN(\sigma).
\end{equation}
It is easy to see that the matrix $e^M$ has all positive entries,
and therefore (by the Perron-Frobenius Theorem) $M$ has a unique
eigenvector $\hat{p}$ of $\ell_1$-norm 1 having all positive
entries. Let $w$ be the eigenvalue corresponding to $\hat{p}$. Then
$w$ is real, it has multiplicity $1$, and it exceeds the real part
of every other eigenvalue. Therefore, for every non-zero vector $y$
with non-negative entries,
\[
\lim_{\tau\to\infty}e^{-\tau w}e^{\tau
M}y=\langle\hat{a},y\rangle\hat{p}
\]
where $\hat{a}$ is the eigenvector of $M^{\trans}$ corresponding to
$w$, normalized so that $\langle\hat{a},\hat{p}\rangle=1$. Note that
$\langle\hat{a},y\rangle>0$ because $y$ is non-zero and
non-negative, and $\hat{a}$ is positive, again by Perron-Frobenius.
Therefore, the vector $\E\left(e^{-\tau w} \cN(\tau)\right)$
converges to a positive scalar multiple of $\hat{p}$, say $\lambda
\hat{p}$, as $\tau \rightarrow \infty$.

In order to prove concentration for the continuous-time model, we
will prove that
the difference $\cN_k(\tau)/q_k- \cN_j(\tau)/q_j $ has an
exponential growth rate which is at most the real part of the second
eigenvalue of $M$, which  is strictly less than $w$, the growth rate
of the individual terms $\cN_k(\tau)/q_k$ and $\cN_j(\tau)/q_j $.
From this, we will conclude that the ratio $\cN_k(\tau)/\cN_j(\tau)$
converges almost surely to $q_k/q_j$, for all $k$ and $j$, which in
turn implies the convergence of the normalized degree sequence to
the vector $(q_i)_{i=0}^{\infty}$.

In order to prove bounds on the growth rate of the differences
$\cN_k(\tau)/q_k- \cN_j(\tau)/q_j $, we will need some auxiliary
bounds involving the
well-known standard birth process, to be defined below.

\subsection{Standard birth process}
\label{app:prclaim}

We start with the definition of the standard birth process with rate $\rho$.
The standard birth process was first introduced by Yule in 1924 \cite{Yule},
and is a special case of the well known Yule Process, defined in that paper.

\begin{defn}
Let $\rho>0$ and let $\{o_n\}_{n=1}^\infty$ be independent exponential random
variables so that $\E(o_n)=\frac{1}{\rho}n^{-1}$. For
$\tau\in[0,\infty)$, let
 $X_{\tau}=\min\{n\geq 1:\sum_{k=1}^no_k>\tau\}$.
Then $X$ is called the {\bf standard birth process with rate $\rho$}.%
\footnote{The name ``standard birth process'' is due to the fact
that $X_\tau$ is equivalent to the following process: Start with one
cell at time $0$. At each time, every cell divides into two cells
with rate $\rho$. Then $X_{\tau}$ is the number of cells at time
$\tau$.}
\end{defn}

The standard birth process is connected to our discussion through
the following easy claim:
\begin{clm}\label{claim:couple}
Let $\|\cN(\tau)\|=\sum_{k=1}^A\cN_k(\tau)$. Let $T\geq 0$, let
$x\geq y$, and let $X$ be a standard birth process with rate $2$. Then
$\left\{\left.\{X_{\tau}\}_{\tau \geq T}\right|X_T=x\right\}$
stochastically dominates $\left\{\left.\{\|\cN(\tau)\|\}_{\tau \geq
T}\right|\|\cN(T)\|=y\right\}$.
\end{clm}

\begin{proof}
Let us start with the observation that $\sum_{k=1}^n o_k$ is the
first time $\tau$ for which $X_\tau=n+1$. Let
$\{r_n\}_{n=0}^{\infty}$ be i.i.d. exponential random variables with
mean 1.  Then $\sum_{k=1}^n o_k$ has the same distribution as
$\sum_{k=0}^{n-1} r_k/(2k+2)$. The time
$\tau_n$ at which the
node $n$ is born has the same distribution as
$\sum_{k=0}^{n-1} r_k/W(k)$, where $W(k)$ denotes the combined
attractiveness of all nodes
at the random time $\tau_k$. The claim follows now from the
observation that
$W(k) \leq 2k+1\leq 2k+2$.
\end{proof}

The main purpose of this section is the proof of the following
claims.

\begin{clm}\label{claim:spenc}
Let $X$ be a standard birth process with rate $\rho$. Then $X_{\tau}$ is
almost surely finite for every $\tau$. Furthermore,
there exists a constant $C_s=C_s(\rho)$ such that for every
$\tau_2>{\tau}_1,$ $x\geq 1,$ and $k\geq 1$,
\begin{equation}\label{eq:spenc}
\prob\left(\left.X_{{\tau}_2}>kxe^{\rho(\tau_2-{\tau}_1)}
\right|X_{{\tau}_1}=x\right) <\frac{C_s}{x(k-1)^2}.
\end{equation}
If, in addition, $\tau_2-\tau_1<1$, then
\begin{equation}\label{eq:spenc2}
\prob\left(\left.X_{{\tau}_2}-X_{{\tau}_1}>kx
[e^{\rho(\tau_2-{\tau}_1)}-1]\right|X_{{\tau}_1}=x\right)
<\frac{C_s}{x(\tau_2-\tau_1)(k-1)^2}.
\end{equation}
\end{clm}

To see the finiteness of $X_{\tau}$, we need to show that
$\sum_{n=1}^\infty o_n=\infty$ a.s. This follows from the following
simple argument: For every $k$,
Let
\[
U_k=\sum_{j=2^k+1}^{2^{k+1}}o_j.
\]
For $j\in[2^k+1,2^{k+1}]$, with probability greater than
$\frac 1 2$, $o_j>\frac 1 \rho 2^{-k-2}$.
Therefore, $\prob(U_k>\frac{1}{4\rho})>\frac 1 2$.
The random variables $\{U_k\}_{k=1}^\infty$ are independent,
and therefore
$\sum_{n=1}^\infty o_n\geq \sum_{k=1}^\infty U_k=\infty$ almost surely.

To see (\ref{eq:spenc}) and (\ref{eq:spenc2}), we use the following
Lemma, which is proved in section II of \cite{Yule}.
Since the proof is short and
simple, we choose to include it for the sake of
making the exposition more self-contained.

\begin{lemma}[Yule, 1924]
\label{lem:toch} For every $\tau>0$ and every positive integer $k$,
$ \E(X_{\tau}^k)<\infty. $ Furthermore,
\begin{equation}\label{eq:expbp}
\E(X_\tau)=\exp(\rho\tau),
\end{equation}
and
\begin{equation}\label{eq:evbp}
\var(X_\tau)=\exp(2\rho\tau)-\exp(\rho\tau).
\end{equation}
In particular,
\begin{equation}\label{eq:varbp}
\var(X_\tau)=O(\exp(2\rho\tau)),
\end{equation}
and for $\tau<1$ there exists a constant $C_v=C_v(\rho)$ so that
\begin{equation}\label{eq:varbpsmall}
\var(X_\tau)\leq
C_v\tau.
\end{equation}
\end{lemma}
\begin{proof}

An equivalent description of the standard birth process is the
following: Let $\alpha$ be an exponential variable with expected
value $\rho$, and
let $\{G_t\}$
be a Poisson point process with
rate $\alpha e^{\rho t}$. Then
$X_\tau=1+G_\tau$
has the same
distribution as the standard birth process. To see this, all we need
is to show that for every $\tau$ and $n$, the rate of the process
$\{G_t\}$ at time $\tau$ conditioned on $X_\tau=n$ is
$\rho n$. Indeed,
\begin{eqnarray*}
\rate(\tau|X_\tau=n)
&=&
\frac{\int_0^\infty
\alpha e^{{\rho}\tau}\prob(X_\tau=n|\alpha)
\frac 1{\rho}
e^{-\alpha/{\rho}}d\alpha}
{\int_0^\infty
\prob(X_\tau=n|\alpha)
\frac 1{\rho}
e^{-\alpha/{\rho}}d\alpha}\\
&=&
\frac{\int_0^\infty
\alpha e^{{\rho}\tau} e^{-\frac\alpha {\rho}(\exp({\rho}\tau)-1)}
\left(\frac\alpha {\rho}(\exp({\rho}\tau)-1)\right)^{(n-1)}
((n-1)!)^{-1}
\frac 1{\rho}
e^{-\alpha/{\rho}}d\alpha}
{\int_0^\infty
e^{-\frac\alpha {\rho}(\exp({\rho}\tau)-1)}
\left(\frac\alpha {\rho}(\exp({\rho}\tau)-1)\right)^{(n-1)}
((n-1)!)^{-1}
\frac 1{\rho}
e^{-\alpha/{\rho}}d\alpha}\\
&=&
{\rho}n.
\end{eqnarray*}
Here the
second equality follows from the fact that
$X_\tau-1$ is a Poisson variable with rate
$\frac\alpha {\rho} (e^{{\rho}\tau}-1)$,
and
last equality follows by integration by parts.

From this we get that the distribution of $X_\tau$ is geometric with
expected value $\exp(\rho\tau)$. To see this,
we again use the fact
that $X_\tau-1$ is a
Poisson variable with rate $\frac\alpha \rho (e^{\rho\tau}-1)$ where
$\alpha$ is an exponential variable with expectation $\rho$. Therefore,
for every $n$,
\begin{eqnarray*}
\prob(X_\tau=n+1)
&=&
\int_0^\infty
\prob(X_\tau-1=n|\alpha)
\frac 1{\rho}
e^{-\alpha/{\rho}}d\alpha\\
&=&
(n!)^{-1}\int_0^\infty
e^{-\frac\alpha {\rho}
(e^{{\rho}\tau}-1)}\left(\frac\alpha {\rho}
(e^{{\rho}\tau}-1)\right)^n
\frac 1{\rho}
e^{-\alpha/{\rho}}d\alpha
=\left(1-e^{-{\rho}\tau}\right)\prob(X_\tau=n)
\end{eqnarray*}
where, again, the last step follows from integration by parts.

The relations
(\ref{eq:expbp}) and (\ref{eq:evbp}) follow immediately, and
(\ref{eq:varbp}) and (\ref{eq:varbpsmall}) follow from
(\ref{eq:evbp}).
\end{proof}

\begin{proof}[Proof of (\ref{eq:spenc}) and (\ref{eq:spenc2})
in Claim \ref{claim:spenc}]
Equations (\ref{eq:spenc}) and  (\ref{eq:spenc2}) will follow from
Chebyshev's inequality if we show that
\begin{equation}\label{eq:et1}
\E(X_{{\tau}_2}|X_{{\tau}_1})=X_{{\tau}_1}e^{\rho(\tau_2-{\tau}_1)}
\end{equation}
and
\begin{equation}\label{eq:vt1}
\var(X_{{\tau}_2}|X_{{\tau}_1})=X_{{\tau}_1}O
\left(e^{2\rho(\tau_2-{\tau}_1)}\right)
\end{equation}
for $\tau_2>{\tau}_1$, and
\begin{equation}\label{eq:jo}
\var(X_{\tau+\tau_1}|X_{\tau_1})=O(\tau)\cdot X_{\tau_1}
\end{equation}
for $\tau<1$.

Equations (\ref{eq:et1}), (\ref{eq:vt1}) and (\ref{eq:jo}) follow
from (respectively) (\ref{eq:expbp}), (\ref{eq:varbp}) and
(\ref{eq:varbpsmall}) and the fact that conditioned on $X_{\tau_1}$,
the process $X_{\tau+\tau_1}$ is the sum of $X_{\tau_1}$ independent
copies of $X_\tau$.
\end{proof}

\noindent
{\bf Remark:} From now on we will always assume that $\rho=2$. In particular,
whenever we use the term "standard birth process", it should be understood
as "standard birth process with rate $2$".

\subsection{Concentration of the continuous-time process}
In order to show concentration of the degree distribution for the
continuous-time process, we will prove first the following lemma. To
state it, we observe for any $b$ with $b^{\trans}\hat{p}=0$,
\begin{equation}\label{eq:rdk1}
\| b^{\trans} e^{(T-\tau)M} \|_{\infty} \leq \|b\|_{\infty}
e^{(T-\tau)v^\prime}
\end{equation}
for some $v^\prime<w$. Without loss of generality, we may assume
that $v^\prime>w/2$. Also, for a general vector $b$,
\begin{equation}\label{eq:rdk2}
\| b^{\trans} e^{(T-\tau)M} \|_{\infty} \leq \|b\|_{\infty}
e^{(T-\tau)w}  .
\end{equation}
\begin{lemma}\label{lemma:mati}
Let $b$ be a vector in $\R^A$ with $\|b\|_{\infty} \leq 1$.  Then
there exists a constant $C<\infty$, such that for all $T>0$,
\begin{equation}\label{eq:secmom}
\var\left(b^{\trans}\cN(T)\right)<C\exp(2u T)
\end{equation}
where $u=w$ if $b^{\trans} \hat p\neq 0$, and $u=v'$ if
$b^{\trans}\hat{p}=0$.
\end{lemma}

\begin{proof}
We use a martingale to bound the variance. Fix $T$, and let
\[L_{\tau}=\E\left(\left.b^{\trans}\cN(T)\right|\cN(\tau)\right). \]
Clearly, $L_{\tau}$ is a (continuous-time) martingale.
By (\ref{eq:exval}), we know that $L_{\tau} = b^{\trans}
e^{(T-\tau)M}\cN(\tau)$. Let $0<\epsilon<\exp(-10T)$ be such that
$K=T/\epsilon$ is an integer number. Then,
$\{U_k=L_{k\epsilon}\}_{k=0}^{K}$ is a martingale and
\[
\var\left(b^{\trans}\cN(T)\right)=\sum_{k=0}^{K-1}\var(U_{k+1}-U_k).
\]
We want to estimate the variance of $U_{k+1}-U_k$. Let $v_k=\cN
((k+1)\epsilon)-\cN (k\epsilon) $. For two vectors $\cN_1$ and
$\cN_2$,
\[
\left(b^{\trans}e^{(T-(k+1)\epsilon)M}\cN_1-b^{\trans}
e^{(T-(k+1)\epsilon)M}\cN_2\right)^2
\leq\|\cN_1-\cN_2\|^2e^{2u(T-(k+1)\epsilon)},
\]
where the norm $\|\cdot\|$ refers to the $L^1$--norm here and
throughout this section, unless otherwise noted. Choose
$\cN(k\epsilon)$ according to its distribution, and let $\cN_1$ and
$\cN_2$ be chosen independently, according to the distribution of
$\cN((k+1)\epsilon)$ conditioned on $\cN(k\epsilon)$. Then
\begin{eqnarray*}
\var(U_{k+1}-U_k)=
\frac12\E\left[\left(b^{\trans}e^{(T-(k+1)\epsilon)M}\cN_1
-b^{\trans}e^{(T-(k+1)\epsilon)M}\cN_2\right)^2\right]
\leq\frac12\E(\|\cN_1-\cN_2\|^2) e^{2u(T-(k+1)\epsilon)}.
\end{eqnarray*}
On the other hand, using the fact that for every vector $x$ in
$\R^d$,
\[
\left(\sum_{i=1}^dx_i\right)^2\leq d\sum_{i=1}^dx_i^2,
\]
we get
\[
\sum_{j=1}^A\var(v_k(j))\geq\frac1{2A}\E(\|\cN_1-\cN_2\|^2)
\]
where $v_k(j)$ is the $j$-th component of $v_k$. Therefore,
\begin{equation}\label{eq:varwithv}
\var(U_{k+1}-U_k)\leq
A\sum_{j=1}^A\exp\left[2u(T-(k+1)\epsilon)\right]\var(v_k(j)).
\end{equation}
By Claim \ref{claim:couple}, (\ref{eq:et1}), and (\ref{eq:jo}), for every
$j=1,2,\ldots,A$,
\[
\var\left(v_k(j)\left|\cN(k\epsilon)\right.\right)
\leq\E\left[(v_k(j))^2\left|\cN(k\epsilon)\right.\right] \leq
C_v\epsilon\|\cN(k\epsilon)\|,
\]
\[
\left|{\E\left(v_k(j)\left|\cN(k\epsilon)\right.\right)}\right|\leq
(e^{2\epsilon}-1)\|\cN(k\epsilon)\| \leq 4\epsilon\|\cN(k\epsilon)\|
\]
and
\[
\E\left[(\|\cN(k\epsilon)\|)^2\right]<e^{4k\epsilon}.
\]
Therefore,
\begin{eqnarray*}
\var(v_k(j))&=&\E\left(\var\left(v_k(j)\left|\cN (k\epsilon)
\right.\right)\right)+
\var\left(\E\left(v_k(j)\left|\cN (k\epsilon)\right.\right)\right)\\
&\leq&
C_v\epsilon\exp(wk\epsilon)+16\epsilon^2\exp(4k\epsilon) \\
&<&
C_0\epsilon\exp(wk\epsilon)
\end{eqnarray*}
for $C_0=C_v+1$, by the choice of $\epsilon$. Therefore,
\begin{eqnarray*}
\var\left(b^{\trans}\cN_k(T)\right) &<&
A^2C_0\epsilon\sum_{k=0}^{K-1}\exp\left(
wk\epsilon+2u(T-(k+1)\epsilon)
\right)\\
&\leq& A^2C_0e^{2u T}\int_0^T e^{(w-2u)\tau}d{\tau} <C_u \exp(2u T)
\end{eqnarray*}
for
\[
C_u = A^2C_0\int_0^\infty e^{(w-2u)\tau}d{\tau}<\infty.
\]
\end{proof}

In addition, note that by (\ref{eq:rdk1}) and (\ref{eq:rdk2}),
\[
\left|\E\left(b^{\trans}\cN(T)\right)\right| \leq e^{uT}
\]
and therefore there exists $C$ so that
\begin{equation}\label{eq:rdk3}
\E\left[(b^{\trans}\cN(T))^2\right] \leq Ce^{2uT}.
\end{equation}

We are now ready to state and prove the two main lemmas used to
prove concentration:

\begin{lemma}\label{lem:egrow}
For every $\wpr<w$ and every $1\leq k\leq A$, a.s. for every $\tau$
large enough,
\begin{equation}\label{eq:gadeldu}
\cN_k(\tau)>e^{\wpr\tau}.
\end{equation}
\end{lemma}
and
\begin{lemma}\label{lem:difnotgrow}
There exists $v<w$ s.t. for every $1\leq k<j\leq A$ a.s. for every
$\tau$ large enough,
\[
p_j\cN_k(\tau)-p_k\cN_j(\tau)<e^{v\tau},
\]
where $p_i, i=1,\ldots,A$ are the components of the vector $\widehat{p}$.
\end{lemma}

The following corollary is an immediate consequence of
Claim~\ref{claim:spenc}, Claim~\ref{claim:couple} and
Lemma~\ref{lem:egrow}.
\begin{cor}
$ w \leq 2.$
\end{cor}

\begin{proof}[Proof of Lemma \ref{lem:difnotgrow}]

Choose some $v$ strictly between $v^\prime$ and $w$  in a way that
$w-v<0.25\min(0.1,v-v^\prime,w/10)$ and let
$\delta=\min(0.1,v-v^\prime,w/10)$. The vector
\[
b_i=\left\{
\begin{array}{ll}
p_j & \mbox{if } i=k \\
-p_k & \mbox{if } i=j \\
0 & \mbox{otherwise}
\end{array}
\right.
\]
satisfies $b^{\trans}\hat p=0$, and therefore, using (\ref{eq:rdk3})
and Markov's inequality,
\begin{equation}\label{eq:frcheb}
\prob\left(p_j\cN_k(T)-p_k\cN_j(T)=b^\trans\cN(T)
>\frac{1}{3}e^{vT}\right)\leq
9Ce^{-2\delta T}.
\end{equation}
Let $\{T_i\}_{i=1,2,\ldots}$ be such that $e^{2\delta T_i}=i^2$. By
Borel-Cantelli, almost surely there exists $i_0$ such that for all
$i>i_0$,
\begin{equation}\label{eq:somt}
p_j\cN_k(T_i)-p_k\cN_j(T_i)<\frac{1}{2}e^{vT_i}.
\end{equation}
Note that
\[
T_i=\frac{\log i}{\delta}
\]
and therefore
\begin{equation}\label{eq:defti}
T_{i+1}-T_i=\Theta(i^{-1}).
\end{equation}
We want to show that almost surely for all $T$ large enough,
\begin{equation}\label{eq:alw}
p_j\cN_k(T)-p_k\cN_j(T)<e^{vT} .
\end{equation}
Section~\ref{subseq:evolexp} tells us that $\E(\| \cN(T_i)
\|)=O(\exp(wT_i))$, and
Lemma \ref{lemma:mati} tells us that
$\var(\| \cN(T_i)\|)=O(\exp(2wT_i))$.   Therefore
\[
\prob(\| \cN(T_i)\|>e^{(w+0.6\delta)T_i})<C_le^{-1.2\delta
T_i}=C_li^{-1.2}
\]
for some constant $C_l$, so that, if $m(i)$ is the number of
vertices arriving between $T_i$ and $T_{i+1}$, then
\begin{eqnarray}\label{eq:befbade}
\nonumber
\lefteqn{\prob\left(m(i)>\frac{1}{2}e^{vT_i}\right)}\\
\nonumber &\leq&\prob\left(\| \cN(T_i)\|>e^{(w+0.6\delta)T_i}\right)
+\prob\left(\left.m(i)>\frac{1}{2}e^{vT_i}\right|\|\cN(T_i)\|
\leq e^{(w+0.6\delta)T_i}\right)\\
\nonumber &\leq&\prob\left(\| \cN(T_i)\|>e^{(w+0.6\delta)T_i}\right)
+\prob\left(\left.m(i)>\frac{1}{2}e^{vT_i}\right|\|\cN(T_i)\|
= e^{(w+0.6\delta)T_i}\right)\\
&\leq& C_li^{-1.2}+C_se^{-(w+0.6\delta)T_i}(T_{i+1}-T_i)^{-1},
\end{eqnarray}
where the last inequality uses (\ref{eq:spenc2}) in Claim
\ref{claim:spenc} and the fact that
\[
\frac{1}{2} e^{vT_i}> 2e^{(w+0.6\delta)T_i}(\exp(2(T_{i+1}-T_i))-1)
\]
for $i$ large enough.

Clearly, the first part of the right side of (\ref{eq:befbade}) is a
convergent sum. We need to show that so is the second part. Remember
the choice $\delta\leq w/10$. Then, using (\ref{eq:defti}),
\begin{eqnarray*}
C_se^{-(w+0.6\delta)T_i}(T_{i+1}-T_i)^{-1}
&=&\Theta\left(i\cdot e^{-(w+0.6\delta)T_i}\right)\\
=\Theta\left(e^{\delta T_i}\cdot e^{-(w+0.6\delta)T_i}\right)
&=&\Theta\left(e^{-(w-0.4\delta)T_i}\right)\\
=O\left(e^{-9\delta T_i}\right) &=&O(i^{-9}).
\end{eqnarray*}
Using
Borel-Cantelli, we conclude that almost surely,
\begin{equation}\label{eq:bade}
\sum_{k=1}^{A}\left|\cN_k(T)-\cN_k(T_i)\right|<\frac{1}{2}e^{{vT_i}}
\end{equation}
for all $k$ and all $i$ large enough and all $T$ between $T_i$ and
$T_{i+1}$. Equation (\ref{eq:alw}) follows from (\ref{eq:bade}).
\end{proof}

\begin{proof}[Proof of Lemma \ref{lem:egrow}]
By Lemma \ref{lemma:mati}, $\var(\cN_1(\tau))<C_1e^{2w\tau}$, while
$\E(\cN_1(\tau))>C_2e^{w\tau}$
by Section~\ref{subseq:evolexp}.
Therefore there exists $\rho>0$ such
that
\begin{equation}\label{eq:usecmom}
\prob\left(\cN_1(\tau)>\rho e^{w\tau}\right)>\rho.
\end{equation}
Fix some large $T$, and let $\tau_i=iT$. For each vertex $v$ which
is a fertile leaf at time $\tau_{i-1}$, let $\ell_v$ denote the
number of descendants of $v$ (including $v$ itself) at time $\tau_i$
which are fertile leaves. The random variables $\{\ell_v\}$ are independent,
their sum
is $\cN_1(\tau_i)$, and the distribution of each of them is the same
as the unconditional distribution of $\cN_1(T)$. Using this fact and
(\ref{eq:usecmom}), we get
\begin{equation}\label{eq:even1}
\prob\left(\left.\cN_1(\tau_i)>\frac{\rho^2}{2}
e^{wT}\cN_1(\tau_{i-1})\right|\cN_1(\tau_{i-1})\right) \geq
1-e^{-\frac{1}{16}\cN_1(\tau_{i-1})}
\end{equation}
via Chernoff's bound. From (\ref{eq:even1}), we get that almost
surely there exists a constant $C_3 > 0$ such that, for all $i$
large enough,
\[
\cN_1(\tau_i)>C_3 \exp\left(i\left[wT+\log\left(\frac{\rho^2}{2}
\right)\right]\right).
\]
From Lemma~\ref{lem:difnotgrow}, we may conclude that the same holds
for $\cN_A(\tau_i)$, i.e. for any constant $C_4 < C_3$,
\[
\cN_A(\tau_i)>C_4 \exp\left(i\left[wT+\log\left(\frac{\rho^2}{2}
\right)\right]\right).
\]
$\cN_A(\tau)$ is monotone increasing, and therefore there exists
$C_5>0$ such that
\begin{equation}\label{eq:befuse}
\cN_A(\tau)>C_5
\exp\left(\tau\left[w+\frac{1}{T}\log\left(\frac{\rho^2}{2}
\right)\right]\right)
\end{equation}
for all $\tau$ large enough. Using Lemma \ref{lem:difnotgrow} again,
we conclude that there exists $C_6>0$ such that
\[
\cN_k(\tau)>C_6
\exp\left(\tau\left[w+\frac{1}{T}\log\left(\frac{\rho^2}{2}
\right)\right]\right)
\]
for all $k$ and large enough $\tau$. We get (\ref{eq:gadeldu}) by
taking $T$ so large that
\[
w+\frac{1}{T}\log\left(\frac{\rho^2}{2}\right)>\wpr.
\]
\end{proof}
\begin{prop}\label{prop:costcont}
For every $k$ and $j$, almost surely
\begin{equation}\label{eq:fuf}
\lim_{t\to\infty}\frac{\cN_k(\tau)}{\cN_j(\tau)}=\frac{p_k}{p_j}
\end{equation}
\end{prop}
\begin{proof}
This follows immediately from Lemma \ref{lem:egrow} and Lemma
\ref{lem:difnotgrow}.
\end{proof}
\subsection{Back to discrete time}
\begin{prop}
For the discrete-time process, and $A>\max\{A_1,A_2\}$ there exists
a vector $\hat{q}$ such that, for $k\leq A$, we have
\begin{equation}
\label{conv-to-qk}
\lim_{\tee\to\infty}\frac{\dN_k(\tee)}{\tee+1}=q_k.
\end{equation}
\end{prop}
\begin{proof}
The number of infertile vertices increases at step $t$ with probability
\[
\frac {\sum_{k=1}^{A_1-1}\dN_k(\tee)} {\sum_{k=1}^{A}\dN_k(\tee)}
\]
(their number cannot decrease). However, by (\ref{eq:fuf}), this expression
tends to a limit, and
therefore, using the law of large numbers,
\begin{equation}\label{eq:valq0}
\lim_{\tee\to\infty}\frac{ N_0(\tee)}{\tee +1}=q_0=\frac
{\sum_{k=1}^{A_1-1}p_k}
{\sum_{k=1}^{A}p_k} .
\end{equation}
Using (\ref{eq:fuf}) once more, the proposition now follows for
$k\geq 1$ with $q_k=(1-q_0)p_k$.
\end{proof}

Note that the above proposition implies that $q_k$ and hence $p_k$
is independent of $A$ if $A>k$, since the left hand side of
\eqref{conv-to-qk} does not depend on $A$ if $A>k$. So, in
particular, $p_1$ does not depend on $A$.

\section{Power Law With a Cutoff}
\label{sec:powerlaw}

In the previous section, we saw that for every $A>\max\{A_1,A_2\}$,
the limiting proportions up to $A-1$ are $\lambda \hat{p}$ where
$\hat{p}$ is the eigenvector corresponding to the highest eigenvalue
$w$ of the $A$-by-$A$ matrix $M$ defined in Eqn.~\ref{eqn:M}.
Therefore, the components $p_1,p_2,\ldots, p_A$ of the vector
$\hat{p}$ satisfy the equation:
\begin{eqnarray}\label{eq:rec}
wp_i=-\min(i,A_2)p_i + \min(i-1,A_2)p_{i-1} \ \ \ \ \ \ i \geq 2
\end{eqnarray}
where the normalization is determined by $\sum_{i=1}^Ap_i=1$. From
(\ref{eq:rec}) we get that for $i\leq A_2$,
\begin{equation}\label{eq:powreg}
p_i=\left(\prod_{k=2}^i\frac{k-1}{k+w}\right)p_1
\end{equation}
and for $i>A_2$
\begin{equation}\label{eq:exreg}
p_i=\left(\frac{A_2}{A_2+w}\right)^{i-A_2}p_{A_2}
\end{equation}
Clearly, (\ref{eq:exreg}) is exponentially decaying. There are many
ways to see that (\ref{eq:powreg}) behaves like a power law with
degree $1+w$.
Indeed,
\begin{eqnarray}\label{eq:jifa}
\frac{p_i}{p_1}=\left(\prod_{k=2}^i\frac{k-1}{k+w}\right)
&=&\exp\left(\sum_{k=2}^i\log\left(\frac{k-1}{k+w}\right)\right)\\
\nonumber
=\exp\left(\sum_{k=2}^i\left(\frac{-1-w}{k+w}\right)+O(1)\right)
&=&\exp\left((-1-w)\left(\sum_{k=2}^i(k+w)^{-1}\right)+O(1)\right)\\
\nonumber
=\exp\left((-1-w)\left(\sum_{k=2}^ik^{-1}\right)+O(1)\right)
&=&\exp\left((-1-w)\left(\sum_{k=2}^i
\log\left(\frac{k+1}{k}\right)\right)+O(1)\right)\\
\nonumber =\exp\left((-1-w)\log(i/2)+O(1)\right)&=&O(1)i^{-1-w}.
\end{eqnarray}
Note that the constants implicit in the $O(\cdot)$ symbols do not
depend on $A_1$, $A_2$ or $i$, due the fact that $0<w\leq 2$.
Equation (\ref{eq:jifa}) can be stated in the following way:

\begin{prop} \label{prop:power-law}
There exist $0<c<C<\infty$ such that for every $A_1$, $A_2$ and
$i\leq A_2$, if $w=w(A_1,A_2)$ is as in (\ref{eq:rec}), then
\begin{equation} \label{eq:PowerLawForP}
ci^{-1-w}\leq\frac{p_i}{p_1}\leq Ci^{-1-w}.
\end{equation}
\end{prop}
The vector $(q_1,q_2,\ldots,q_{A-1})$ is a scalar multiple of the
vector $(p_1,p_2,\ldots,p_{A-1})$, so equations (\ref{mthm:1}),
(\ref{mthm:2}), and (\ref{eqn:mainthm_powerlaw}) in
Theorem~\ref{thm:main-thm} (and the comment immediately following
it) are consequences of equations (\ref{eq:powreg}),
(\ref{eq:exreg}), and (\ref{eq:PowerLawForP}) derived above.  It
remains to prove the normalization conditions
\[
\sum_{i=0}^{\infty} q_i  =  1 \qquad {\rm and} \qquad
q_0  =  \sum_{i=1}^{\infty} q_i \min(i-1, A_1-1)
\]
stated in Theorem~\ref{thm:main-thm}.  These follow from the
equations
\[
\sum_{i=0}^{\infty} N_i(t)  =  t+1 \qquad {\rm and} \qquad
N_0(t)  =  \sum_{i=1}^{\infty} N_i(t) \min(i-1, A_1-1).
\]
The first of these simply says that there are $t+1$ vertices at time
$t$; the second equation is proved by counting the number of
infertile children of each fertile node.

\section{Monotonicity Properties of $w$}
\label{appdx:monotonicity} In this section we will prove that the
exponent $1+w$ of the power law in Proposition~\ref{prop:power-law}
is monotonically decreasing in $A_1$ and monotonically increasing in
$A_2$.  For this purpose, it will be useful to define a family of
matrices, parameterized by two vectors $\yvec, \zvec \in \R^n$, which
generalizes the matrix $M$ appearing in (\ref{eqn:M}), whose top
eigenvalue is $w$.

Given vectors $\yvec = (y_1,y_2,\ldots,y_n), \zvec =
(z_1,z_2,\ldots,z_n) \in \R^n$, let $\Myz$ denote the $n$-by-$n$
matrix whose $(ij)$-th entry is:
\[
M_{i,j}(\yvec,\zvec) = \left\{
\begin{array}{ll}
z_1 - y_1 & \mbox{if } 1=i=j \\
-y_j & \mbox{if } \ 2 \leq i=j \leq n \\
y_j & \mbox{if } \ 2 \leq i=j+1 \leq n \\
z_j & \mbox{if } \ i=1 \mbox{ and } j \geq 2 \\
0 & \mbox{otherwise}.
\end{array}
\right.
\]
Thus, for instance, the matrix $M$ defined in (\ref{eqn:M}) is
$\Myz$, where $n=A$ and
\begin{eqnarray*}
y_j & = & \left\{ \begin{array}{ll}
\min(j,A_2) & \mbox{if } 1 \leq j < A \\
0 & \mbox{if } j=A
\end{array} \right. \\
z_j & = & \left\{ \begin{array}{ll}
0 & \mbox{if } 1 \leq j < A_1 \\
\min(j,A_2) & \mbox{if } A_1 \leq j \leq A.
\end{array} \right.
\end{eqnarray*}
For the remainder of this section, we will assume:
\begin{eqnarray}
\label{eqn:yi}
& \bullet &  y_i > 0 \mbox{ for } 1 \leq i < n, \\
\label{eqn:zi}
& \bullet &  z_i \geq 0 \mbox{ for } 1 \leq i < n,  \\
\label{eqn:zn} & \bullet & y_n = 0, \; z_n > 0.
\end{eqnarray}
All of these criteria will be satisfied by the matrices $\Myz$ which
arise in proving the desired monotonicity claim.  It follows from
(\ref{eqn:yi}),(\ref{eqn:zi}), and (\ref{eqn:zn}) that if we add a
suitably large scalar multiple of the identity matrix to $\Myz$, we
obtain an irreducible matrix $\Myz + B I$ with non-negative entries.
The Perron-Frobenius Theorem guarantees that $\Myz + BI$ has a
positive real eigenvalue $R$ of multiplicity 1, such that all other
complex eigenvalues have modulus $\leq R$; consequently $\Myz$ has a
real eigenvalue $w = R-B$, of multiplicity 1, such that the real
part of every other eigenvalue is strictly less than $w$.

We will study how $w$ varies under perturbations of the parameters
$\yvec,\zvec$.  Let $P(\lambda,\yvec,\zvec)$ be the characteristic
polynomial of $\Myz$, i.e.
\[
P(\lambda,\yvec,\zvec) = \det(\lambda I - \Myz).
\]
This is a polynomial of degree $n$ in $\lambda$ (with coefficients
depending smoothly on $\yvec,\zvec$), whose largest real root
$w(\yvec,\zvec)$ exists and has multiplicity 1, provided $(\yvec,
\zvec)$ belongs to the region $V \subset \R^n \times \R^n$
determined by (\ref{eqn:yi}),(\ref{eqn:zi}), and (\ref{eqn:zn}).  It
follows from the implicit function theorem that $w(\yvec,\zvec)$ is
a smooth function of $(\yvec,\zvec)$ in $V$, satisfying:
\begin{equation} \label{eqn:IFT}
\left. \left( \frac{\partial P}{\partial y_i} \, + \, \frac{\partial
w}{\partial y_i} \cdot \frac{\partial P}{\partial \lambda} \right)
\right|_{(w,\yvec,\zvec)} = 0; \qquad \left. \left( \frac{\partial
P}{\partial z_i} \, + \, \frac{\partial w}{\partial z_i} \cdot
\frac{\partial P}{\partial \lambda} \right)
\right|_{(w,\yvec,\zvec)} = 0.
\end{equation}
If $\xvec$ is any vector in $\R^n \times \R^n$, and
$\partial_{\xvec}$ is the corresponding directional derivative
operator, we have from (\ref{eqn:IFT}):
\begin{equation} \label{eqn:dirdiv}
\partial_{\xvec} w(\yvec,\zvec) =
- \frac{\partial_{\xvec} P(w,\yvec,\zvec)} {(\partial P / \partial
\lambda)|_{(w,\yvec,\zvec)}}.
\end{equation}
We know that $(\partial P / \partial \lambda)|_{(w,\yvec,\zvec)} >
0$ because $P$ is a polynomial with positive leading coefficient,
$w$ is its largest real root, and $w$ has multiplicity 1.  Thus we
have established:
\begin{clm} For any vector $\xvec \in \R^n \times \R^n$,
and any $(\yvec,\zvec) \in V$, put $w = w(\yvec,\zvec)$.  Then the
directional derivatives $\partial_{\xvec} w(\yvec,\zvec)$ and
$\partial_{\xvec} P(w,\yvec,\zvec)$ have opposite signs.
\label{claim:dirdiv}
\end{clm}
This allows monotonicity properties of $w$ to be deduced from
calculations involving directional derivatives of $P$. Given the
definition of $\Myz$, it is straightforward to compute that
\begin{equation} \label{eqn:charpoly}
P(\lambda,\yvec,\zvec) \; = \; \det(\lambda I - \Myz) \; = \;
P_1(\lambda,y,z) \; -  \; \sum_{j=2}^n P_j(\lambda,\yvec,\zvec),
\end{equation}
where
\begin{eqnarray} \label{eqn:p1}
P_1(\lambda,y,z) & = &
(\lambda+y_1-z_1) \prod_{i=2}^n (\lambda + y_i)  \\
 \label{eqn:pj}
P_j(\lambda,\yvec,\zvec) & = & \left( \prod_{i=1}^{j-1} y_i \right)
z_j \left( \prod_{i=j+1}^n (\lambda + y_j) \right).
\end{eqnarray}
As an easy consequence of this formula, $w$ is strictly positive.
\begin{lemma}
$w$ is strictly positive.
\end{lemma}
\begin{proof}
From (\ref{eqn:charpoly})--(\ref{eqn:pj}) and the fact that $y_n=0$,
we have $P(0,\yvec,\zvec)=-P_n(0,\yvec,\zvec)
=-\left(\prod_{i=1}^{n-1} y_i\right) z_n$, and this is strictly
negative by (\ref{eqn:yi}) and (\ref{eqn:zn}). For sufficiently
large positive $\lambda$, we know that $P(\lambda,\yvec,\zvec)>0$
because $P$ is a polynomial whose leading coefficient in $\lambda$
is positive.  By the intermediate value theorem, $P(\lambda,\yvec,
\zvec)$ has a strictly positive real root.
\end{proof}

The following three lemmas encapsulate the requisite directional
derivative estimates for $P$.
\begin{lemma} \label{lem:mono1}
$(\partial P / \partial z_k)|_{(w,\yvec,\zvec)} < 0$ for $(\yvec,
\zvec) \in V$.
\end{lemma}
\begin{proof}
For $k>1$,
\[
\partial P / \partial z_k =
- \partial P_k / \partial z_k = - \left( \prod_{i=1}^{k-1} y_i
\right) \left( \prod_{i=k+1}^n (w + y_i) \right) < 0.
\]
For $k=1$,
\[
\partial P / \partial z_1 =
\partial P_1 / \partial z_1 =
- \prod_{i=2}^n (w + y_i) < 0.
\]
\end{proof}
\begin{cor}
\label{cor:wA1} $w$ is monotonically decreasing in $A_1$.
\end{cor}
\begin{proof}
Increasing $A_1$ from $k$ to $k+1$ has no effect on $\yvec$, and its
only effect on $\zvec$ is to decrease $z_k$ from $\min(k,A_2)$ to 0.
As we move in the $-z_k$ direction, the directional derivative of
$P$ is positive, so the directional derivative of $w$ is negative by
Claim~\ref{claim:dirdiv}. Thus $w$ decreases as we increase $A_1$
from $k$ to $k+1$.
\end{proof}
\begin{lemma} \label{lem:mono2}
For $1 < k < n$, $(\partial P / \partial y_k)|_{(w,\yvec,\zvec)} <
0$ if $(\yvec,\zvec) \in V$ and $z_k=0$.
\end{lemma}
\begin{proof}
\begin{eqnarray*}
\frac{\partial P}{\partial y_k} & = & \frac{\partial P_1}{\partial
y_k}  \; - \;
\sum_{j=2}^n \frac{\partial P_j}{\partial y_k} \\
& = &
\frac{1}{w+y_k} P_1 \; - \; \frac{1}{w+y_k} \sum_{j=2}^{k-1} P_j \;
- \;
\frac{1}{y_k} \sum_{j=k+1}^n P_j \\
& < &
\frac{1}{w+y_k} P_1 \; - \; \frac{1}{w+y_k} \sum_{j=2}^{k-1} P_j \;
- \;
\frac{1}{w+y_k} \sum_{j=k+1}^n P_j \\
& = & \frac{P(w,\yvec,\zvec)}{w+y_k}  \\
& = & 0
\end{eqnarray*}
\end{proof}
\begin{lemma} \label{lem:mono3}
For $k>1$, $ (\partial P / \partial y_k +
\partial P / \partial z_k)|_{(w,\yvec,\zvec)} < 0$
if $(\yvec,\zvec) \in V$ and $y_k = z_k$.
\end{lemma}
\begin{proof}
\begin{eqnarray*}
\frac{\partial P}{\partial y_k} + \frac{\partial P}{\partial z_k} &
= &
\frac{\partial P_1}{\partial y_k} \; - \; \sum_{j=2}^n
\frac{\partial P_j}{\partial y_k}
\; - \; \frac{\partial P_k}{\partial z_k} \\
& = &
\frac{1}{w+y_k} P_1 \; - \; \frac{1}{w+y_k} \sum_{j=2}^{k-1} P_j \;
- \; \frac{1}{y_k} \sum_{j=k+1}^n P_j \; - \;
\frac{1}{z_k} P_k \\
& < &
\frac{1}{w+y_k} P_1 \; - \; \frac{1}{w+y_k} \sum_{j=2}^{k-1} P_j \;
- \; \frac{1}{w+y_k} \sum_{j=k+1}^n P_j \; - \;
\frac{1}{w+y_k} P_k \\
& = & \frac{P(w,\yvec,\zvec)}{w+y_k}  \\
& = & 0
\end{eqnarray*}
\end{proof}
\begin{cor}
\label{cor:wA2} $w$ is monotonically increasing in $A_2$.
\end{cor}
\begin{proof}
If we change $A_2$ from $k$ to $k+1$, this changes
$\yvec$ into a new vector $\yvec'$ satisfying
\[
y'_j - y_j = \left\{ \begin{array}{ll}
1 & \mbox{if } k < j < n \\
0 & \mbox{otherwise.}
\end{array} \right.
\]
It changes
$\zvec$ into a new vector $\zvec'$ satisfying
\[
z'_j-z_j = \left\{ \begin{array}{ll}
1 & \mbox{if } \max(A_1,k+1) \leq j \leq n \\
0 & \mbox{otherwise.}
\end{array} \right.
\]
Letting $\mathbf{e}_j^{(y)}$ denote a unit vector in the $+y_j$
direction, and $\mathbf{e}_j^{(z)}$ a unit vector in the $+z_j$
direction, the direction of change is expressed by the vector
\[
\xvec = (\yvec',\zvec') - (\yvec,\zvec) = \left[ \sum_{k+1 \le j <
A_1} \mathbf{e}_j^{(y)} \right] + \left[\sum_{\max(k+1,A_1) \le j <
n} (\mathbf{e}_j^{(y)} + \mathbf{e}_j^{(z)}) \right] +
\mathbf{e}_n^{(z)},
\]
and $\partial_{\xvec} P$ is negative, by the preceding three lemmas.
By Claim~\ref{claim:dirdiv}, this means $w$ increases monotonically
as we move along this path.
\end{proof}

\section{Acknowledgment}
We wish to thank the referee for many good suggestions that helped us improve
the paper.

\bibliographystyle{plain}

\end{document}